%% file: manuscript.tex
\documentclass[preprint2]{aastex6}

\usepackage{xspace}

\usepackage{verbatim}
\usepackage[T1]{fontenc}
\usepackage{ifthen}
\usepackage{float}
\usepackage{numprint}
\usepackage{dcolumn,booktabs}
\usepackage{scrextend}

\newcolumntype{d}[1]{D{.}{.}{#1}}

\newcommand{\Kepler}{\textit{Kepler}\xspace}

\newcommand{\Mstar}{\ensuremath{M_{\star}}\xspace}
\newcommand{\Rstar}{\ensuremath{R_{\star}}\xspace} 
 
\newcommand{\fe}{\ensuremath{\mathrm{[Fe/H]}}\xspace}
\newcommand{\teff}{$T_{\mathrm{eff}}$\xspace}  
\newcommand{\logg}{\ensuremath{\log g}\xspace} 
\newcommand{\vsini}{\ensuremath{v \sin i}\xspace}

\newcommand{\Mp}{\ensuremath{M_{P}}\xspace} 
\newcommand{\Mcore}{\ensuremath{M_{\mathrm{core}}}\xspace}
\newcommand{\Menv}{\ensuremath{M_{\mathrm{env}}}\xspace}

\newcommand{\Qprime}{\ensuremath{Q^{\prime}}\xspace}

\newcommand{\Rp}{\ensuremath{R_P}\xspace}
\newcommand{\Teq}{$T_{\mathrm{eq}}$\xspace}

\newcommand{\ms}{\ensuremath{\mathrm{m}\,\mathrm{s}^{-1}}\xspace}
\newcommand{\kms}{\ensuremath{\mathrm{km}\,\mathrm{s}^{-1}}\xspace}
\newcommand{\msyr}{\ensuremath{\mathrm{m}\,\mathrm{s}^{-1}\,\mathrm{yr}^{-1}}\xspace}

\newcommand{\Me}{\ensuremath{M_{\oplus}}\xspace} 
\renewcommand{\Re}{\ensuremath{R_{\oplus}}\xspace} 

\newcommand{\gcc}{g~cm$^{-3}$\xspace}
\newcommand{\Rsun}{\ensuremath{R_{\odot}}\xspace }
\newcommand{\Msun}{\ensuremath{M_{\odot}}\xspace}
 
\newcommand{\rhostar}{\ensuremath{\rho_\star}\xspace}

\newcommand{\bjdtdb}{\ensuremath{\mathrm{BJD}_\mathrm{TBD}}\xspace}
\def\deg{\ensuremath{^{\circ}}}

\newcommand{\lonperi}{\ensuremath{\omega_{}}\xspace}

\newcommand{\esinw}{\ensuremath{e \sin \lonperi}\xspace}
\newcommand{\sqrtecosw}{\ensuremath{\sqrt{e} \cos \lonperi}\xspace}
\newcommand{\sqrtesinw}{\ensuremath{\sqrt{e} \sin \lonperi}\xspace}
\newcommand{\dvdt}{\ensuremath{\dot{\gamma}}\xspace}

\newcommand{\sigjit}[1]{
        \ensuremath{
                \ifthenelse{\equal{#1}{}}{\sigma_\mathrm{jit}}{\sigma_\mathrm{jit,#1}}}
        \xspace
}
\newcommand{\gam}[1]{\ensuremath{\gamma_\mathrm{#1}}\xspace}

\newcommand{\dbic}{\ensuremath{\Delta\mathrm{BIC}}\xspace}
\input{all_params.tex}

\begin{document}


\title{Kepler-1656\MakeLowercase{b}: a Dense Sub-Saturn With an Extreme Eccentricity}



\author{Madison~T.~ Brady\altaffilmark{1}}
\author{Erik A. Petigura\altaffilmark{1,7}}
\author{Heather A. Knutson\altaffilmark{1}}
\author{Evan Sinukoff\altaffilmark{2,3}}
\author{Howard Isaacson\altaffilmark{4}}
\author{Lea~A.~Hirsch\altaffilmark{4}}
\author{Benjamin~J.~Fulton\altaffilmark{1,5}}, 
\author{Molly R. Kosiarek\altaffilmark{6,8}}
\author{Andrew W. Howard\altaffilmark{3}}

\altaffiltext{1}{Division of Geological and Planetary Sciences, California Institute of Technology, Pasadena, CA 91125, USA}
\altaffiltext{2}{Institute for Astronomy, University of Hawai`i at M\={a}noa, Honolulu, HI 96822, USA}
\altaffiltext{3}{Cahill Center for Astrophysics, California Institute of Technology, Pasadena, CA 91125, USA}
\altaffiltext{4}{Department of Astronomy, University of California, Berkeley, CA 94720, USA}
\altaffiltext{5}{NASA Exoplanet Science Institute \& Infrared Processing and Analysis Center, California Institute of Technology,  Pasadena, CA, 91125, USA}
\altaffiltext{6}{University of California Santa Cruz, Santa Cruz, CA, 95064, USA}
\altaffiltext{7}{Hubble Fellow}
\altaffiltext{8}{NSF Graduate Research Fellow}

\begin{abstract}
Kepler-1656b is a $5~\Re$ planet with an orbital period of 32 days initially detected by the prime \Kepler mission. We obtained precision radial velocities of Kepler-1656 with Keck/HIRES in order to confirm the planet and to characterize its mass and orbital eccentricity. With a mass of $48\pm4$~\Me, Kepler-1656b is more massive than most planets of comparable size. Its high mass implies that a significant fraction, roughly 80\%, of the planet's total mass is in high-density material such as rock/iron, with the remaining mass in a low-density H/He envelope. The planet also has a high eccentricity of $0.84\pm0.01$, the largest measured eccentricity for any planet less than 100~\Me. The planet's high density and high eccentricity may be the result of one or more scattering and merger events during or after the dispersal of the protoplanetary disk.
\end{abstract}


\section{Introduction}

Among the eight known planets in the solar system, there are notable gaps in the distribution of planet sizes. There are no planets between the size of Uranus (4.0 $\Re$) and Saturn (9.1 $\Re$). Before the discovery of extrasolar planets, it was unclear whether or not nature produced planets of intermediate sizes.  Today, thanks largely to NASA's {\em Kepler Space Telescope} \citep{Borucki10a}, we know that such planets do exist. 

Planets between the size of Earth and Neptune are ubiquitous. There are about 30 such planets per 100 Sun-like stars with orbital periods less than 100 days \citep{Petigura18b}. While planets between the size of Neptune and Saturn, ``sub-Saturns,'' are roughly 10 times more rare than planets between the size of Earth and Neptune, sub-Saturns offer valuable windows into planet formation physics not accessible among solar system objects. In particular, they offer a valuable test of theories of giant planet formation by core-nucleated accretion (e.g., \citealt{Pollack96}), which were first formulated to explain the solar system planets. Sub-Saturns also provide observational leverage on planet population synthesis models. Prior to \Kepler, such models often predicted that sub-Saturns would be extremely rare (e.g., \citealt{ida:2004}), while more recent models more closely match the \Kepler planet population (e.g., \citealt{Jin14}).

A growing number of exoplanets have well-measured masses and sizes, which reflect their bulk compositions.  Mass measurements are usually made with radial velocities (RVs) or transit-timing variations (TTVs), while planet sizes are constrained with transit photometry. Measurements of planet masses and radii during the prime \Kepler mission revealed an important transition, in the bulk properties of planets: Planets smaller than 1.5 $\Re$ typically have high densities consistent with rocky compositions, while larger planets require substantial envelopes of low-density material, likely H/He \citep{Weiss14, Marcy14, Rogers15}.  

Sub-Saturns offer a different window into the internal structure of planets.  Their large radii imply that envelopes of H/He make up a significant percentage of their overall mass \citep{Lopez14}, with typical envelope fractions ranging from $10-50\%$ \citep{petigura17a}. The distribution of mass between core and envelope provides important clues regarding the formation of these objects.

Here, we report the RV confirmation and characterization of Kepler-1656b, a sub-Saturn identified as a planet candidate during the prime \Kepler mission. We describe our RV follow-up and \Kepler photometric monitoring in Section~\ref{sec:ob}. Section~\ref{sec:an} describes our joint model of the Kepler-1656 RVs and photometry, which revealed a high density and high eccentricity for Kepler-1656b. In Section~\ref{sec:discussion}, we place Kepler-1656b in the context of the broader exoplanet population and consider possible formation scenarios. We conclude in Section~\ref{sec:conclusion}.

\section{Observations}
\label{sec:ob}

\subsection{Spectroscopy}
\label{ssec:spect}

Kepler-1656 (a.k.a KOI-367, KIC-4815520) was observed using the High Resolution Echelle Spectrometer (HIRES; \citealt{vogt:1994}) on the 10m Keck Telescope I.  We collected 100 spectra between 2016-05-12 and 2017-07-11 through an iodine cell mounted directly in front of the spectrometer slit.  This cell imprinted a dense forest of absorption lines to be used as a wavelength reference. An exposure meter was used to achieve a consistent signal-to-noise level of 110 per reduced pixel on blaze near 550 nm. We also obtained a ``template'' spectrum without the use of the iodine cell.

Our RVs were determined using standard California Planet Search procedures \citep{howard:2010}.  These include forward modeling of the stellar and iodine spectra convolved with the instrumental response \citep{Marcy92, Valenti95}.  The measurement uncertainty of each point ranged from about 1.5 to 2.5 \ms and was derived from the uncertainty on the mean RV of the approximately 700 spectral chunks used in the RV pipeline.  

We also measured the Mount Wilson $S_\mathrm{HK}$ activity index, which traces the chromospheric emission in the cores of the Ca II HK lines \citep{vaughan78}.  The $S_\mathrm{HK}$ index is a tracer of stellar activity, which produces apparent RV variability in some stars. Kepler-1656 has a median $S_\mathrm{HK}$ of 0.15, which is similar to that of other low-activity stars of similar $B-V$ color \citep{Isaacson10}. Table \ref{tab:rv} lists our RV and $S_\mathrm{HK}$ measurements.

Early in our observational campaign, we noticed that RVs taken on the same night frequently had an RMS dispersion of around $4~\ms$, larger than the formal measurement uncertainties.  Kepler-1656 may be just beginning to evolve off the main sequence, which could account for its additional, short timescale RV variability, i.e. ``jitter.''  Whatever the cause, we obtained three exposures separated by an hour whenever possible to average over stellar variability having timescales less than 1 hr. This is similar to noise mitigation observing strategy recommended by \cite{Dumusque11}.

\begin{deluxetable}{rrrrr}
\tablecaption{Radial Velocity and Activity Measurements\label{tab:rv}}
\tablecolumns{5}
\tablewidth{-0pt}
\tabletypesize{\footnotesize}
\tablehead{
        \colhead{Time} & 
        \colhead{RV} & 
        \colhead{$\sigma$(RV)}&
        \colhead{$S_\mathrm{HK}$} &
        \colhead{$\sigma (S_\mathrm{HK})$} \\
        \colhead{\bjdtdb} & 
        \colhead{\ms} & 
        \colhead{\ms} &
        \colhead{} &
        \colhead{}
        }
\decimals
\startdata
\input{tab_rv_stub.tex}
\enddata
\tablecomments{\ref{tab:rv} is published in its entirety in machine-readable format. A portion is shown here for guidance regarding its form and content.}
\end{deluxetable}

\subsection{Photometry}
\label{ssec:photo}

Kepler-1656 was observed by \Kepler during its prime mission (2009--2013; \citealt{Borucki10a}). We downloaded median-detrended long-cadence \Kepler photometry ($\sim$49000 measurements) from the NASA Exoplanet Archive \citep{Akeson13}.%
\footnote{https://exoplanetarchive.ipac.caltech.edu/}

\subsection{Imaging}
\label{ssec:imaging}
The Exoplanet Follow-up Observing Program  (ExoFOP) archive%
\footnote{https://exofop.ipac.caltech.edu/}
contains several observations of Kepler-1656 by different high-resolution imaging facilities, which place limits on the presence of stellar companions. The strongest constraints come from the Differential Speckle Survey Instrument (DSSI) on the 8m Gemini-N telescope \citep{Furlan17}. No source with contrast $\Delta i < 4.4$~mag was found down to separations of 100~mas. Consulting the Gaia DR2 parallax \citep{Gaia18} and the empirical mean stellar color sequence from \cite{pecaut:2013}, we found that the DSSI observations excluded physically bound companions earlier than M4 at projected separations larger than 18 au. These limits are relevant to understanding the origin of Kepler-1656b's extreme eccentricity and are discussed in Section~\ref{sec:discussion}.

\section{Analysis}
\label{sec:an}

When a planet has a circular orbit, the RV time series and photometric time series constrain separate planetary properties; thus, they can be modeled independently without any loss of generality.  However, in the case of eccentric orbits, both the RV time series and the transit profile constrain both the orbital eccentricity $e$ and argument of periastron $\omega$.  We first model RVs independently (Section~\ref{ssec:rv}). We then motivate the inclusion of photometry in our orbit modeling (Section~\ref{ssec:tr}) and present our joint model (Section~\ref{ssec:joint}).

\subsection{RV Model}
\label{ssec:rv}
We modeled our RVs using the publicly available RV-modeling Python code \texttt{RadVel} \citep{Fulton18}. We did not bin RV measurements taken on the same night. Our single planet model included the following parameters: orbital period $P$, transit time $T_0$, orbital eccentricity $e$, longitude of periastron $\omega$, Doppler semi-amplitude $K$, RV offset $\gamma$, and RV acceleration $\dvdt$.  We also included RV jitter \sigjit{} that accounts for RV variability due to non-planet sources (such as instrumental noise and stellar variability) into our likelihood function:

\begin{equation}
\label{eqn:rvlike}
    \ln{\mathcal{L}} = 
        - \frac{1}{2}\sum_{i} 
        \left[
            \frac{\left(v_{\mathrm{obs},i}-v_{\mathrm{mod},i}\right)^2}{\sigma_{\mathrm{obs}, i}^2+ \sigjit{}^2}
            + \ln{2\pi\left(\sigma_{\mathrm{obs},i}^2 + \sigjit{}^2\right)}
        \right],
\end{equation}

\begin{figure*}
\centering
\includegraphics[width=0.7\textwidth]{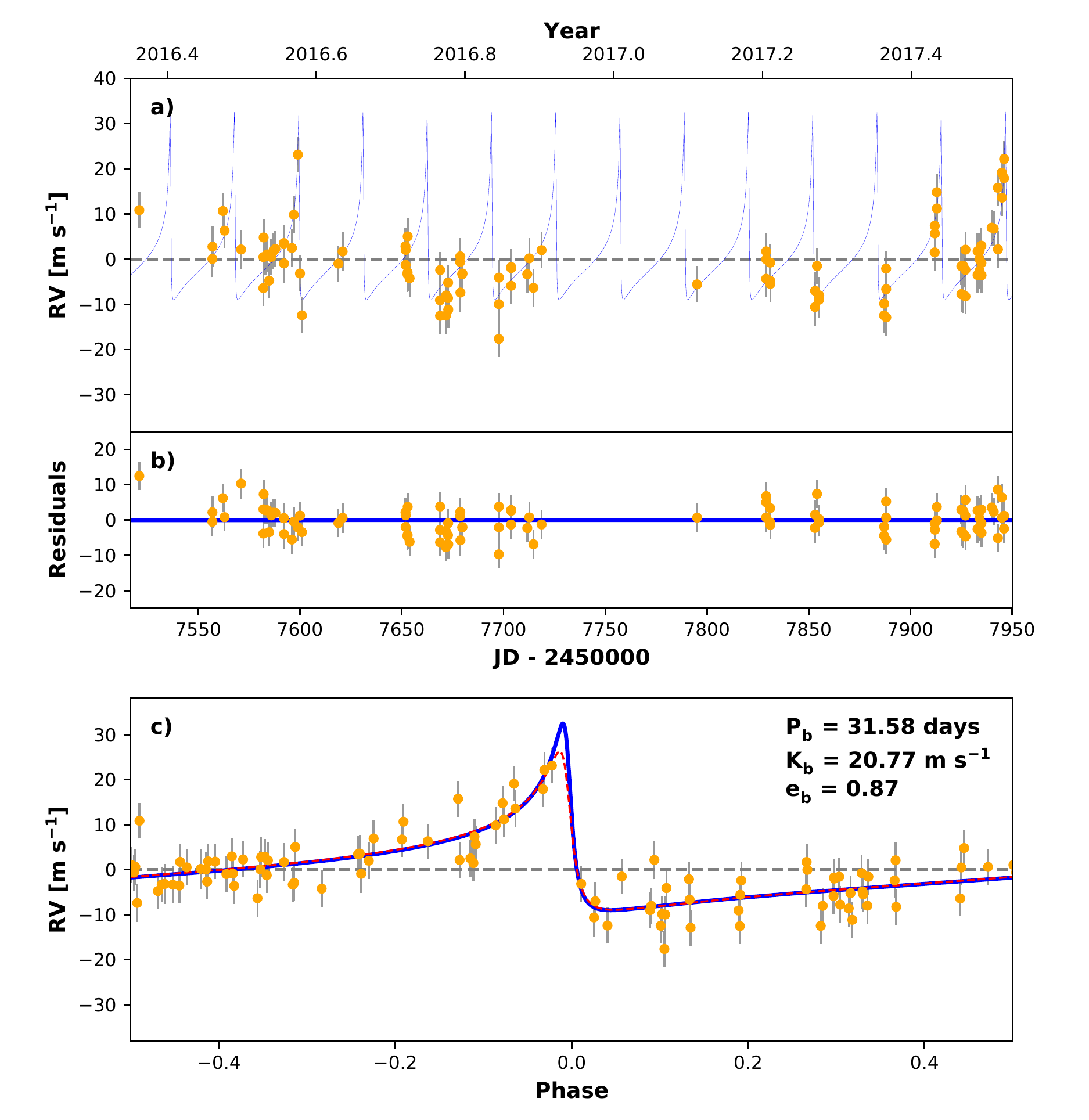}
 \caption{Single Keplerian model of Kepler-1656 RV data with nonzero eccentricity (see Section~\ref{ssec:rv}). {\bf a)} Time series of RVs from HIRES. The blue line shows the most probable Keplerian model. {\bf b)} Residuals to the most probable Keplerian model. {\bf c)} Phase-folded RVs of the most probable Keplerian. The red dashed line shows the most probable Keplerian from the final adopted model described in Section~\ref{ssec:joint}. \label{fig:rvmcmc}}
\end{figure*}

We first considered whether inclusion of a nonzero eccentricity was motivated by the data by using the Bayesian Information Criterion (BIC; \citealt{schwarz:1978}).  We performed a maximum-likelihood fit for both eccentric and circular models.  The BIC strongly favors eccentric over circular models with $\dbic = -50$.  Next, we considered whether the data motivated a nonzero \dvdt, which would be indicative of the existence of an additional long-period companion.  Including \dvdt was not favored, with $\dbic = +5$.  The most probable RV model is shown in Figure~\ref{fig:rvmcmc}.  We also checked for possible correlations between RV and stellar activity. After subtracting the most probable Keplerian model, we found that our model residuals had no significant correlation with the $S_\mathrm{HK}$ index, with a Pearson $r^2$ coefficient of $0.04$.

We used \texttt{RadVel's} MCMC tools to compute uncertainties on our RV parameters. \texttt{RadVel} automatically tests for convergence using the Gelman--Rubin statistic \cite{Gelman03}. The 1$\sigma$ credible range on our model parameters are listed in Table~\ref{tab:koi-367}. The RVs indicate a high eccentricity of $e = \val{KOI-367-rv}{ecc}$.  Even though the BIC favored $\dvdt = 0$, we included it in the final version of our fits in order to quantify an upper limit of $\dot{\gamma} < 1.84$ \msyr. 

\subsection{Transits and Orbital Eccentricity}
\label{ssec:tr}

\begin{figure*}
\centering
\includegraphics[width=0.8\textwidth]{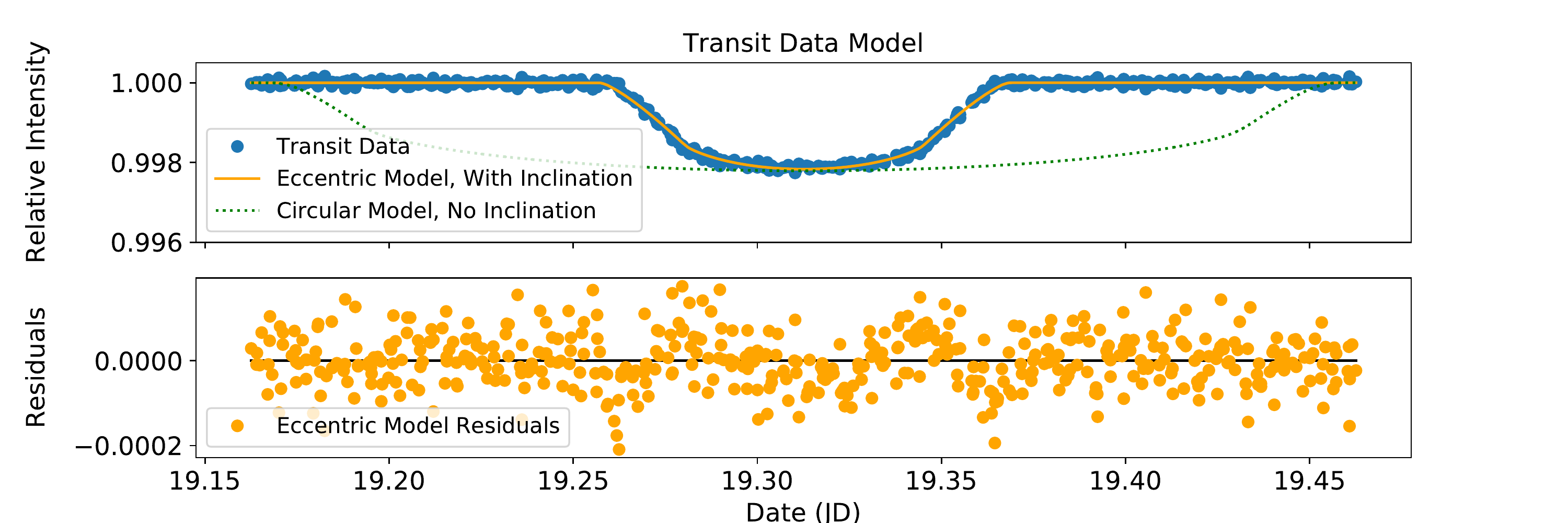}
\caption{Top panel: the phase-folded \textit{Kepler} photometry data, with our best-fit transit model shown in yellow.  Details of the fitting are discussed in Section~\ref{ssec:joint}. A model of the transit of a $90^{\circ}$ planet on a circular orbit is included for reference. Bottom panel: residuals to our most-probable photometry model.  \label{fig:trmcmc}}
\end{figure*}

Transits are often used to constrain a planet's orbital period, planet-to-star radius ratio $\Rp / \Rstar$, and time of conjunction $T_c$. However, the transit duration also encodes eccentricity information. A planet on a circular orbit has a constant velocity
\[
    v_{\mathrm{circ}} = \sqrt{\frac{G\Mstar}{a}}.
\]
The maximum transit duration (assuming an impact parameter of $b = 0$) is given by
\[
    T_{\mathrm{14}} = \frac{\Rstar P}{\pi a}.
\]
For eccentric orbits, the orbital speed is not constant, and thus the duration of a central transit depends on eccentricity and viewing geometry \citep{seager:2003,Winn10}:
\[
    T_{\mathrm{14}} = \frac{\Rstar P}{\pi a} \frac{\sqrt{1 - e^2}}{1 \pm \esinw}.
\]
The relationship between transit duration and eccentricity is often referred to as the  ``photo-eccentric effect'' (e.g. \citealt{dawson:2012}). It has been used to constrain the eccentricities of individual planets and large ensembles of planets (see, e.g., \citealt{vaneylen15} and \citealt{Xie16}). Transits provide the most eccentricity information when the following conditions are met:

\begin{enumerate}
\item \textit{The mean stellar density is known precisely.}
For a planet on a circular orbit, $T_{14} \propto \Rstar / a \propto \rhostar^{-1/3}$. Therefore, one is most sensitive to deviations from circular orbits when mean stellar density is known precisely. For Kepler-1656b, \rhostar is measured to about 30\% (see Table~\ref{tab:koi-367}). 

\item \textit{The light curve has a high signal-to-noise ratio.} 
A high signal-to-noise ratio is necessary to resolve degeneracies between the transit duration and impact parameter.  Figure \ref{fig:trmcmc} shows that ingress/egress are resolved in the transit profile of Kepler-1656b.

\item \textit{The transit duration is different from that of a circular orbit.}  
If the planet transits near periapse or apoapse, its duration will differ significantly from the value expected for a circular orbit. For Kepler-1656b, the transit duration is one-third that of a centrally transiting planet with a circular orbit (see Figure~\ref{fig:trmcmc}). This was an early indication that Kepler-1656b may be a high-eccentricity object with transit occurring near periapse.
\end{enumerate}

Given that these criteria were met for Kepler-1656b, we decided to perform a joint modeling of the RVs and photometry.

\subsection{Joint Model}
\label{ssec:joint}
We modeled the transit profile using the publicly available Python package \texttt{batman} \citep{Kreidberg15}.  We assumed normally distributed photometric errors and adopted the following likelihood:
\begin{equation}
\label{eqn:transitlike}
    \ln{\mathcal{L}} = 
        - \frac{1}{2}\sum_{i} 
            \frac{(f_{\mathrm{obs},i}-f_{\mathrm{mod},i})^2}{\sigma_{\mathrm{obs},i}^2}.
\end{equation}
Here, $f_{\mathrm{obs},i}$ is the relative flux measured at time $t_\mathrm{i}$, $\sigma_{\mathrm{obs},i}$ is the uncertainty of this measurement, and $f_{\mathrm{mod},i}$ is the calculated model flux at $t_i$. We adopted a quadratic limb-darkening model, adopting values of $u = 0.4160$ and $v = 0.2496$ as computed by \cite{sing09} for a $T_{\mathrm{eff}} = 5750$ K and  $\logg = 4.50$ star.  We found that the modeled uncertainties in the limb-darkening coefficients had a negligible effect on our results. 

Because the RV and photometric datasets are independent, our joint likelihood is the sum of Equations \ref{eqn:rvlike} and \ref{eqn:transitlike}. The joint likelihood is function of the following 10 free parameters: $K$, $\gamma$, \dvdt, $T_0$, $a/\Rstar$, $\Rp/\Rstar$, $\cos{i}$, \sqrtecosw, and \sqrtesinw. We fixed the period to $P = 31.578659$ days, as reported in the Exoplanet Archive.  We imposed a Gaussian prior on $a/\Rstar$ based on spectroscopic measurements by the  California-\Kepler Survey \citep{Petigura17b,Johnson17}. 

We explored the likelihood surface with Markov Chain Monte Carlo (MCMC) , using the affine-invariant sampler of \cite{Goodman10}. We initialized 20 walkers and ran the chains for $N_\mathrm{step} =  10^6$ steps, discarding the first $10^5$ as burn in. For each parameter, we computed the autocorrelation time $\tau$ and found that it was at most $2\times10^4$ steps. Given that $\tau \ll N_{step}$, we concluded that the chains are well-mixed. We include a corner plot that highlights the covariance that exists between some parameters in the Appendix. This covariance is strongest for \sqrtecosw and \sqrtesinw.

\begin{figure}
\includegraphics[width=0.48\textwidth]{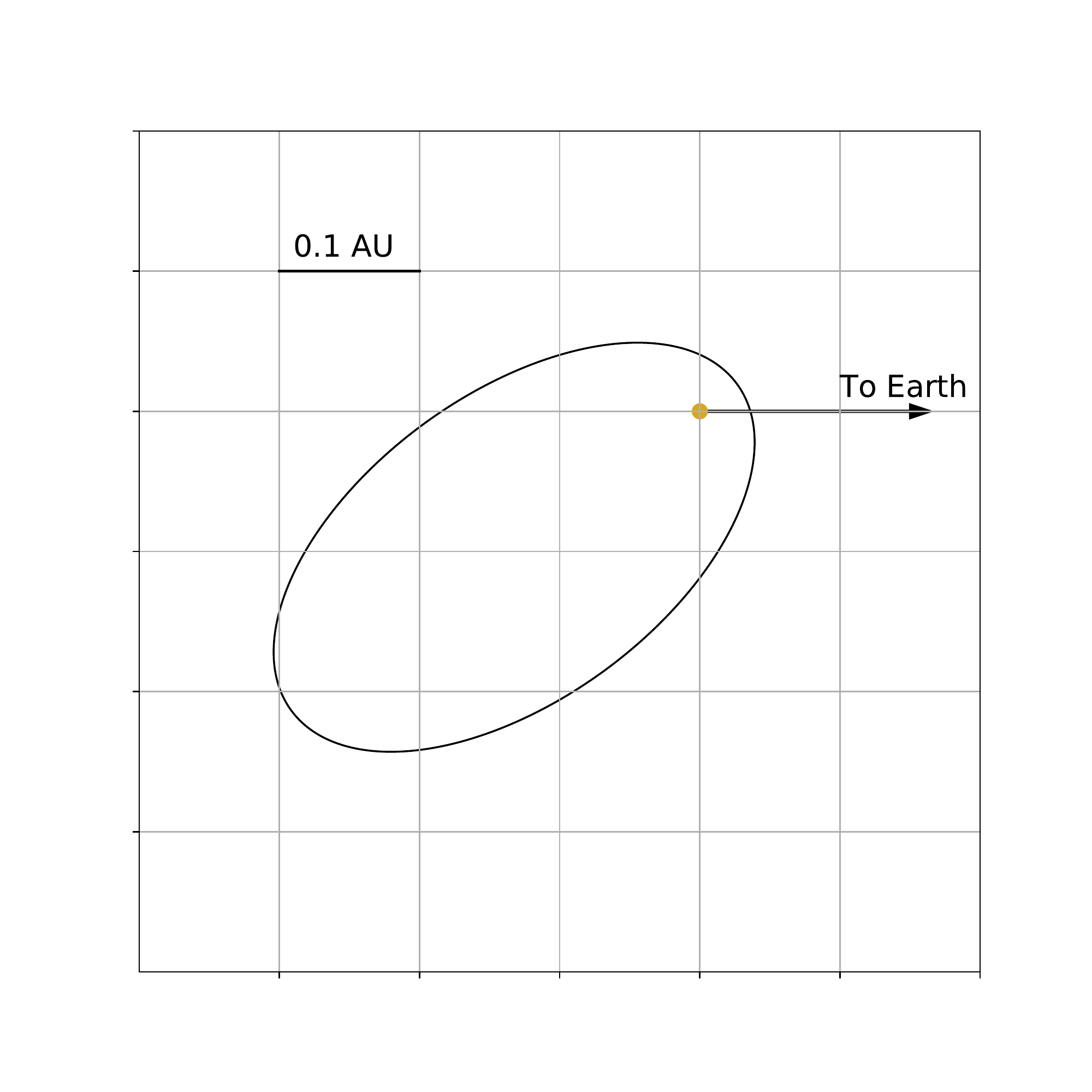}
\caption{The eccentric orbit of Kepler-1656b. Star is drawn to scale. \label{fig:orbmodel}}
\end{figure}

The 1$\sigma$ credible ranges on our model parameters are listed in Table~\ref{tab:koi-367}. Using our joint model, we found an eccentricity of $e = \val{KOI-367-jo}{ecc}$. This value is consistent with the RV-only model, but has uncertainties that are four times smaller. The extra precision stems from the complementary constraints from transits and RVs. Figure~\ref{fig:orbmodel} shows the planet's eccentric orbit relative to its host star. We discuss formation scenarios that can account for this high eccentricity in Section~\ref{sec:discussion}.

{\renewcommand{\arraystretch}{0.9}
\begin{table}
\begin{center}
\caption{System parameters of Kepler-1656}
\begin{tabular}{lrr}
\hline
\hline
Parameter              & Value   & Notes \\
\multicolumn{3}{l}{{\bf Stellar parameters}} \\
\teff (K)           & $5731 \pm 60$   & A \\
\logg (dex)         & $4.37 \pm 0.10$ & A \\
\fe (dex)           & $0.19 \pm 0.04$ & A \\
\vsini (\kms)       & $2.8  \pm 1.0$  & A \\
\Mstar (\Msun)      & $1.03\pm 0.04$  & B \\
\Rstar (\Rsun)      & $1.10 \pm 0.13$ & B \\
age (Gyr)           & $\val{KOI-367-star}{agegyr}$ & B \\
$V$ (mag)           & $\val{KOI-367-star}{vmag}$   & C \\
$K$ (mag)           & $\val{KOI-367-star}{kmag}$   & D \\
\multicolumn{3}{l}{{\bf RV-Only Model}} \\
$P$ (days)          & $\val{KOI-367-rv}{P1}$ (fixed)  & E\\
$T_0$ (BJD)         & $\val{KOI-367-rv}{T01}$ (fixed) & E \\
$K$ (\ms)           & $\val{KOI-367-rv}{k}$           & E \\
\gam{} (\ms)        & $\val{KOI-367-rv}{gamma}$       & E \\
\dvdt (\msyr)       & $\val{KOI-367-rv}{dvdt}$        & E \\
\sigjit{} (\ms)     & $\val{KOI-367-rv}{jit}$         & E \\
$e$                 & $\val{KOI-367-rv}{ecc}$         & E \\
$\omega$ (deg)      & $\val{KOI-367-rv}{omega}$       & E \\
\\[-2ex]
\multicolumn{3}{l}{{\bf Adopted Joint Model}} \\
$P$ (days)          & $\val{KOI-367-rv}{P1}$ (fixed)  & E \\
$T_0$ (BJD)         & $\val{KOI-367-jo}{T01}$  & E \\
$K$ (\ms)           & $\val{KOI-367-jo}{k}$           & E \\
\gam{} (\ms)        & $\val{KOI-367-jo}{gamma}$       & E \\
\dvdt (\msyr)       & $\val{KOI-367-jo}{dvdt}$        & E \\
\sigjit (\ms)       & $\val{KOI-367-jo}{jit}$         & E \\
$a / \Rstar$        & $\val{KOI-367-jo}{a_r}$         & E \\
$\Rp / \Rstar$      & $\val{KOI-367-jo}{rp_r}$        & E \\
$i$ (deg)           & $\val{KOI-367-jo}{inc}$         & E \\  
$e$                 & $\val{KOI-367-jo}{ecc}$         & E \\
$\omega$ (deg)      & $\val{KOI-367-jo}{omega}$       & E \\
\\[-2ex]
\multicolumn{3}{l}{{\bf Derived Parameters}} \\
\Mp (\Me)           & $\val{KOI-367-jo}{mp}$          & E \\
$a$ (au)            & $\val{KOI-367-jo}{a}$           & E \\
$\Rp$ (\Re)         & $\val{KOI-367-jo}{rp}$          & E \\
$\rho$ (\gcc)       & $\val{KOI-367-jo}{dp}$          & E \\
\Teq (K)            & $\val{KOI-367-jo}{tp}$          & E \\
$b$                 & $\val{KOI-367-jo}{b}$           & E \\
\\[-2ex]       
\hline
\\[-6ex]       
\end{tabular}
\end{center}
\tablecomments{A: \cite{Petigura17b}. B: \cite{Johnson17}. C: \cite{hog:2000}. D: \cite{skrutskie:2006}. E: This work.}
\label{tab:koi-367}
\end{table}
}

\section{Discussion}
\label{sec:discussion}
\subsection{Mass, Radius, and Envelope Fraction}
\label{ssec:mass-and-radius}

\begin{figure*}
\includegraphics[width=0.5\linewidth]{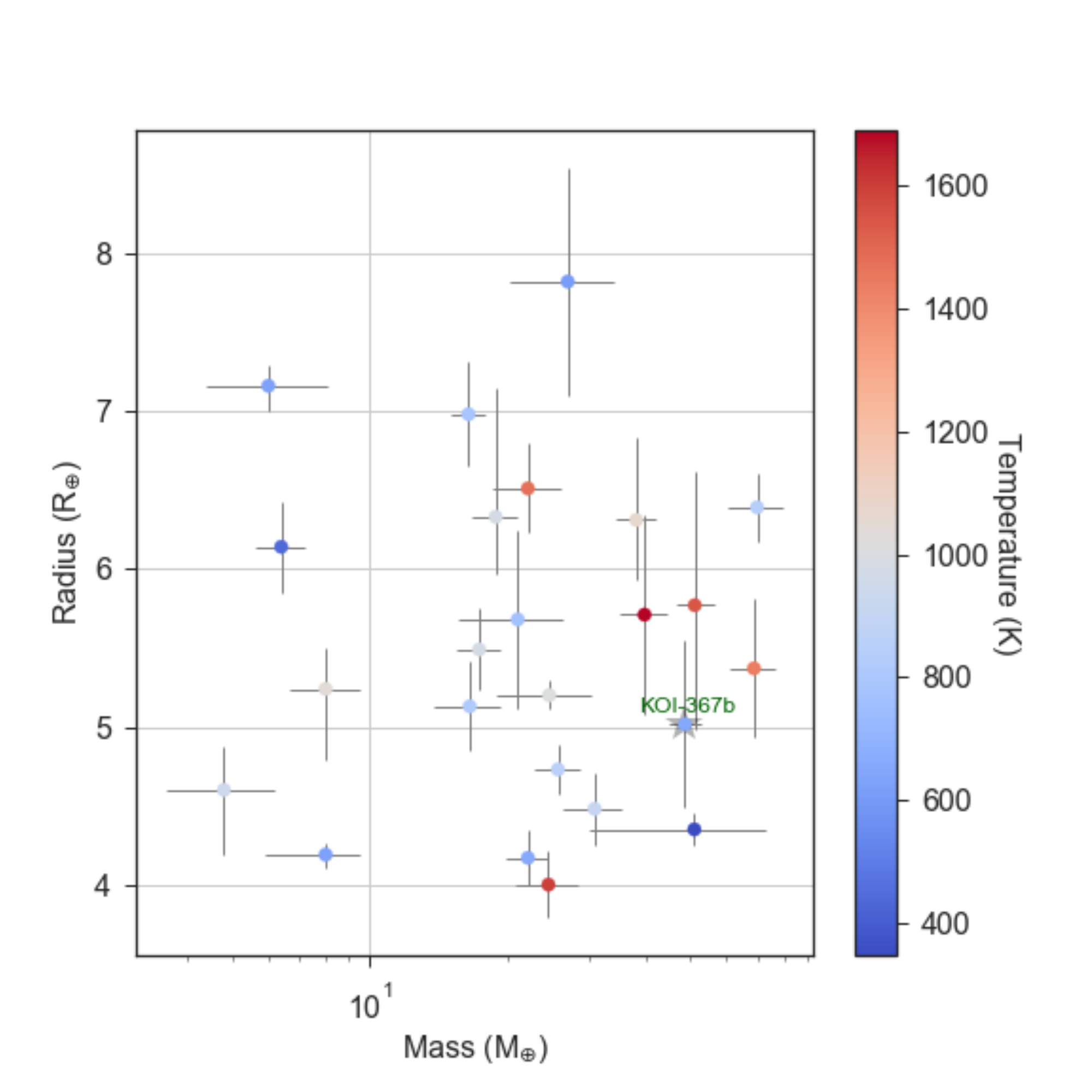}
\includegraphics[width=0.5\linewidth]{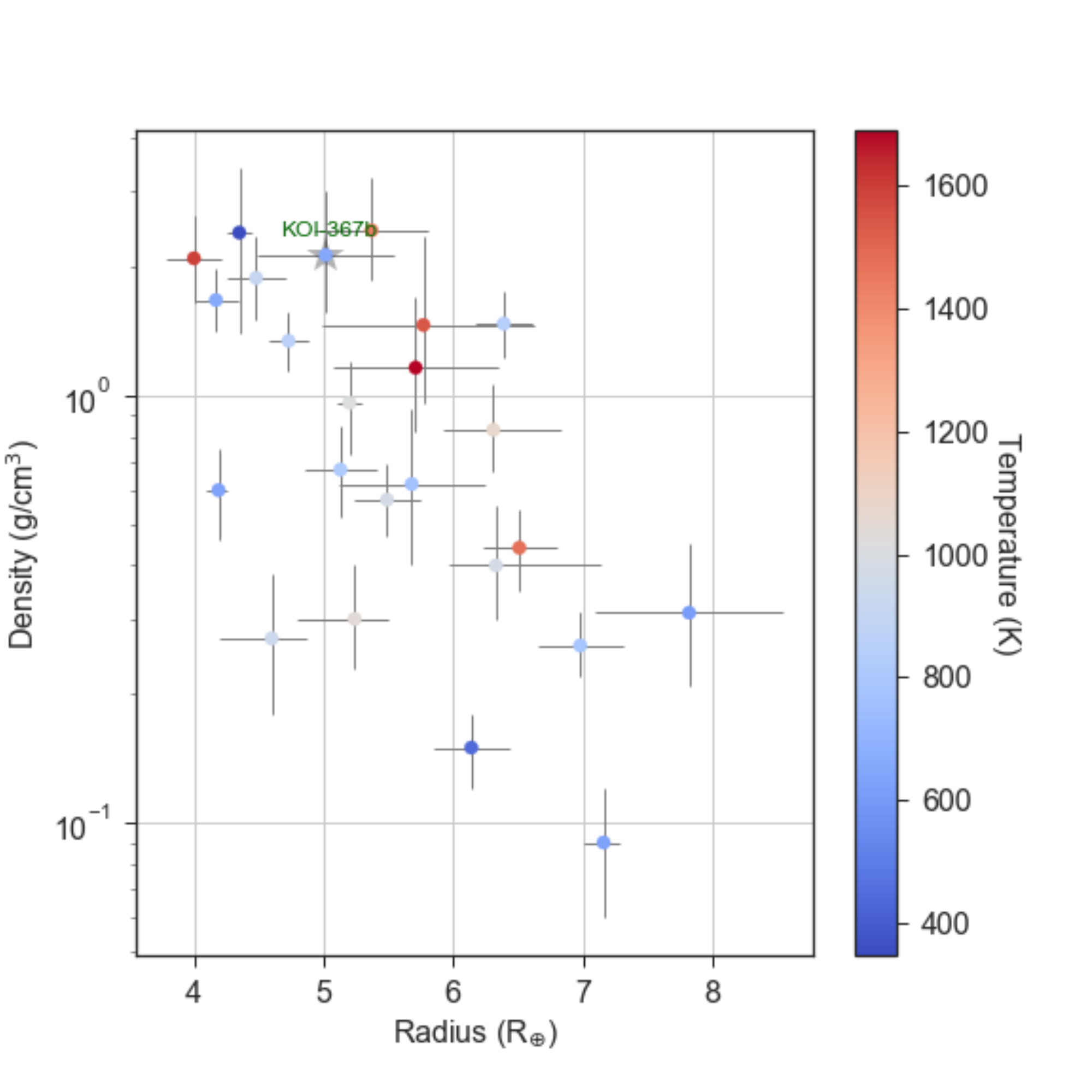}
\caption{Left panel: the mass and radius of Kepler-1656b in the context of other sub-Saturns from \cite{petigura17a}. All points are colored according to planetary blackbody equilibrium temperature. Right panel: same as left, but showing density and radius. Kepler-1656b is among the most massive and highest density sub-Saturns known.}
\label{fig:rp_mp_rhop}
\end{figure*}

Kepler-1656b is a member of a class of planets between the size of Neptune and Saturn or ``sub-Saturns.'' \cite{petigura17a} studied a sample of 20 sub-Saturns and observed an order of magnitude dispersion of density at a given size; this indicated a diversity in the core-envelope structure of these planets. Figure~\ref{fig:rp_mp_rhop} shows the mass, radius, and density of Kepler-1656b in the context of other sub-Saturns with well-measured sizes from the \cite{petigura17a} sample. Kepler-1656b has a mass of $\val{KOI-367-jo}{mp}~\Me$, making it one of the most massive sub-Saturns known. Its high mass also implies a high density, which at $\rho = \val{KOI-367-jo}{dp}$ \gcc makes Kepler-1656b one of the densest sub-Saturns known and denser than any gaseous object in the solar system. 

Following \cite{petigura17a}, we quantified the core and envelope fractions of Kepler-1656b using the \cite{Lopez14} planet structure models, which assume an Earth composition core and a envelope of primordial H/He. In the sub-Saturn size range, derived envelope fractions are not sensitive to the precise composition of the core \citep{Petigura16}. Under these assumptions, we found that $82\pm6\%$ of the planet's mass is in the core. This value is on the high end of what is observed among sub-Saturns.

\subsection{Eccentricity}
\label{ecc}
A planet's present-day eccentricity is a relic of its formation and migration history. However, there are many plausible channels for exciting eccentricity. For a review of different excitation mechanisms, see \cite{Dawson18} and references therein. Over the past twenty years, RV surveys have found that giant exoplanets often have much higher eccentricities than those of solar system planets \citep{Winn15}. 
 
\begin{figure*}
\centering
\includegraphics[width=1.0\textwidth]{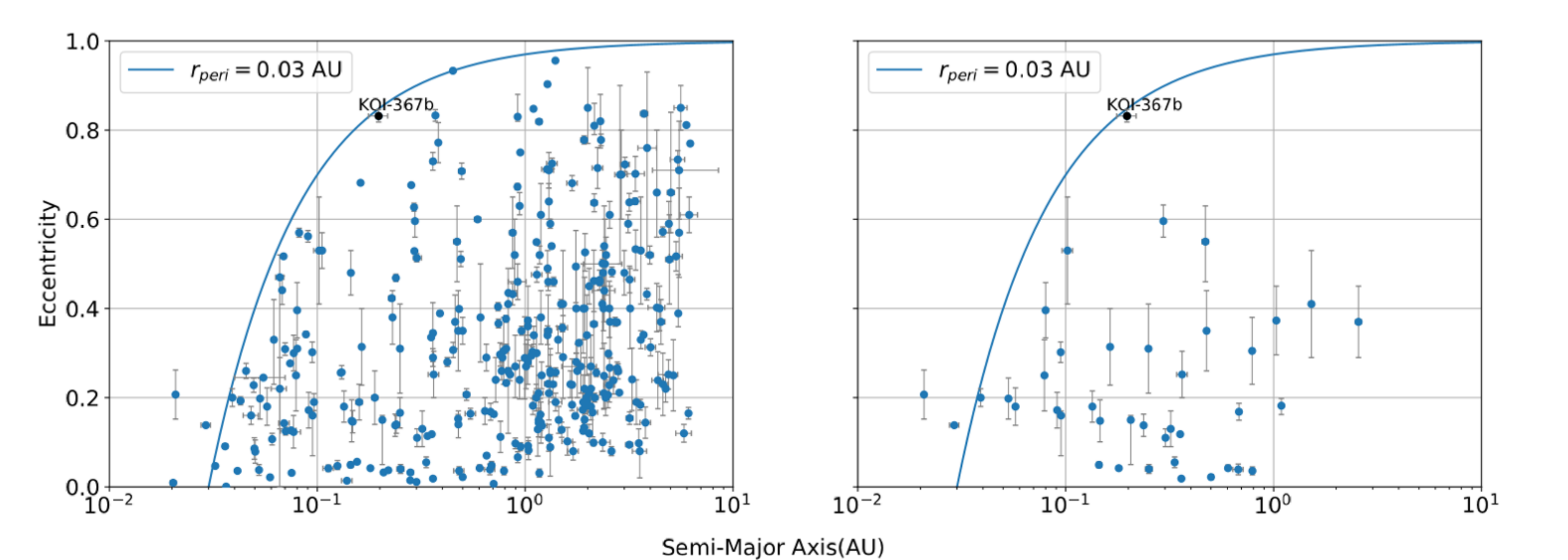}
\caption{Left panel: the eccentricity and semi-major axis of confirmed exoplanets with well-measured eccentricities (source: NASA Exoplanet Archive). The line indicates orbits where  $r_{peri} = 0.03$ au. Right panel: same as left, but only showing planets where $< 100 \Me$. \label{fig:eccsma}}
\end{figure*}

Figure~\ref{fig:eccsma} shows the eccentricity and semi-major axis for all planets where eccentricity is known to better than $2\sigma$ (NASA Exoplanet Archive; \citealt{Akeson13}). With an eccentricity of 0.84, Kepler-1656b is one of the most eccentric planets known.

The upper envelope of the $e$--$a$ distribution can be approximated by orbits with periastron distance of 0.03~au, and is likely due to rapid tidal circularization of planets whose orbits take them within $0.03$~au of their host stars. For low eccentricity orbits, \cite{Goldreich66} showed that the timescale for eccentricity damping, $\tau_e$, is given by:
\begin{equation}
\label{eqn:taue}
    \tau_e = \frac{4}{63} 
                 \left(\frac{\Qprime}{n}\right) 
                 \left(\frac{\Mp}{\Mstar}\right)
                 \left(\frac{a}{\Rp}\right)^5.
\end{equation}
Here, $n = \sqrt{G \Mstar / a^3}$ is the mean motion, and \Qprime, the modified tidal quality factor, is given by $\Qprime = 3 Q / 2 k_2$, where $Q$ is the specific dissipation function and $k_2$ is the Love number. The steep dependence on $a$ naturally leads to a well-defined upper envelope in the $e$--$a$ plane, although the details depend on $Q^{\prime}$, which is uncertain at the order of magnitude level.

Given that the periastron of Kepler-1656b is currently near this critical value of 0.03~au, there is a possibility that it is undergoing tidal circularization. Assessing this possibility requires applying tidal theory at high-eccentricity (e.g., \citealt{Hut82}) and will be treated in a subsequent work.

Most of the exoplanets with well-measured eccentricities have Jovian masses, due to larger Doppler signals. Eccentricity   There are only  $\sim30$ planets with well-measured eccentricities where $\Mp < 100 \Me$. Among these, Kepler-1656b has by far the highest eccentricity. At these low masses, the $e$--$a$ distribution of planets is uncertain due to the limited number of measurements. For example, it is unclear whether the $r_\mathrm{peri} = 0.03$~au envelope also applies to this low-mass population.

It is possible that the formation pathway that creates giant planets favors high eccentricities compared to the processes that produce the more common super-Earths and sub-Neptunes.  Transit duration analyses of \Kepler planets find low typical eccentricities for systems with multiple transiting planets  $\langle e\rangle$ = 0.04, while systems with single transiting planets show a broader range of typical eccentricities $\langle e\rangle \approx 0.30$ \citep{Xie16}. Such transit duration studies are statistical in nature, and do not directly probe any connections between planet mass and eccentricity. Additional eccentricity measurements of low-mass planets are needed.

\subsection{Formation}
\label{ssec:add_plt}

With a core mass of roughly 40~\Me, Kepler-1656b presents challenges to the classical theory of giant planet formation by core-nucleated accretion. In the canonical models of \cite{Pollack96}, gas giants form from rocky cores that are on the order of 10~\Me. These cores initially accrete gas slowly at a rate that is limited by radiative cooling of the envelope. When $\Mcore\sim\Menv$, runaway accretion sets in, and the planet quickly accretes all the gas in its feeding zone. Somehow the core of Kepler-1656b grew to roughly 40~\Me, but never underwent runaway accretion. 

One possible resolution is that the core of Kepler-1656b reached its final mass during or after the disk dispersal stage. For instance, Kepler-1656b could be the result of the merger of several sub-critical cores. Viscous interactions with the gas disk would have sufficiently damped the growth of eccentricity of these cores during the disk phase. As the disk dispersed, their orbital eccentricities would have grown, eventually resulting in an orbit crossing and merger.  After this merger, Kepler-1656b would then have accreted its modest envelope from the depleted gas disk.

Alternatively, this merger might have taken place after the gas disk was fully dispersed.  In this scenario, the precursor objects would have had their own gas envelopes.  Simulations of merging gaseous planets have demonstrated that, under certain conditions, mergers can concentrate solids by preferentially disrupting the more lightly bound gaseous envelopes \citep{Liu15b}.

Planet-planet scattering resulting in a merger could also help to explain the high eccentricity of Kepler-1656b. Such a merger, however, is not the only outcome of a scattering event. Planet-planet scattering can also lead to the ejection of one of the planets. However, this is unlikely because Kepler-1656b is deep in its host star's gravitational well. The maximum velocity Kepler-1656b can impart onto another planet is its surface escape velocity \citep{goldreich:2004}, which is 35~\kms. Because the surface escape velocity is smaller than the orbital velocity of approximately 70~\kms, Kepler-1656b cannot impart a large enough velocity kick to eject objects. Therefore, the most likely outcome of a planet-planet scattering is a merger.

Although a merger between planets could potentially explain this planet's high orbital eccentricity, this might also result from secular interactions with another body in the system. However, our upper limit of $\dot{\gamma} < 1.4$ \msyr implies that objects of comparable mass to Kepler-1656b have $a \gtrsim 4$~au. Furthermore, archival high-resolution imaging rules out bound stellar companions earlier than M4, with $a \gtrsim 18$~au (Section~\ref{ssec:imaging}). While the existing observations cannot rule out all stellar and planetary companions that could excite eccentricity via secular interactions, we consider such mechanisms unlikely.

We also considered stellar flybys as a possible mechanism to excite Kepler-1656b's eccentricity. \cite{Li15} computed the cross section for eccentricity pumping interactions due to passing stars (Equation~19 in their paper). They found that such interactions are far more likely to occur in a star's birth cluster compared to the field. However, for Kepler-1656b, the interaction cross section is quite small due to its small semi-major axis. For nominal cluster parameters, the probability that a stellar flyby would appreciably change Kepler-1656b's eccentricity is $\sim 10^{-3}$ and is therefore an unlikely explanation for the planet's eccentric orbit.

\section{Conclusions}
\label{sec:conclusion}
In this work, we combined transit photometry from \Kepler and RVs from Keck/HIRES to measure the mass, radius, and eccentricity of the sub-Saturn Kepler-1656b. The planet is massive compared to other planets of comparable size, which suggests that a large fraction of the planet's mass is in a high-density core. We also found that Kepler-1656b has an extreme eccentricity, the highest known for a planet with <100~\Me.

Kepler-1656b may be the product of one or more merger events during or after the stage of disk dispersal. These events worked to concentrate a large mass of solids, excite eccentricity, and clear the system of other planets on neighboring orbits. If this story is correct, its formation pathway would have been very different than that of the solar system planets.

Of course, it is difficult to piece together the detailed formation history of any individual exoplanet given the limited number of observables, but a strength of exoplanet astronomy is the ability to probe many outcomes of planet formation physics in different systems. Additional measurements of masses, radii, eccentricities, and other properties of sub-Saturns will shed additional light on a class of planets not represented in the solar system. 

\clearpage
\acknowledgments 
We thank Konstantin Batygin for useful discussions, as well as Lauren Weiss and Ian Crossfield for their assistance with Keck/HIRES observations.

M.T.B. acknowledges support from a Summer Undergraduate Research Fellowship (SURF) at Caltech, which was funded with the assistance of Hannah Bradley. E.A.P. acknowledges support from a Hubble Fellowship grant HST-HF2-51365.001-A awarded by the Space Telescope Science Institute, which is operated by the Association of Universities for Research in Astronomy, Inc. for NASA under contract NAS 5-26555. 

The data presented herein were obtained at the W. M. Keck Observatory, which is operated as a scientific partnership among the California Institute of Technology, the University of California and the National Aeronautics and Space Administration. The Observatory was made possible by the generous financial support of the W. M. Keck Foundation.
The authors wish to recognize and acknowledge the very significant cultural role and reverence that the summit of Maunakea has long had within the indigenous Hawaiian community.  We are most fortunate to have the opportunity to conduct observations from this mountain.

\software{Astropy \citep{Astropy13}, batman \citep{Kreidberg15}, emcee \citep{Foreman-Mackey13},
Numpy/Scipy \citep{VanDerWalt11}, Matplotlib \citep{Hunter07}, Pandas \citep{McKinney10},  RadVel \citep{Fulton18}}

\begin{figure*}
\gridline{\fig{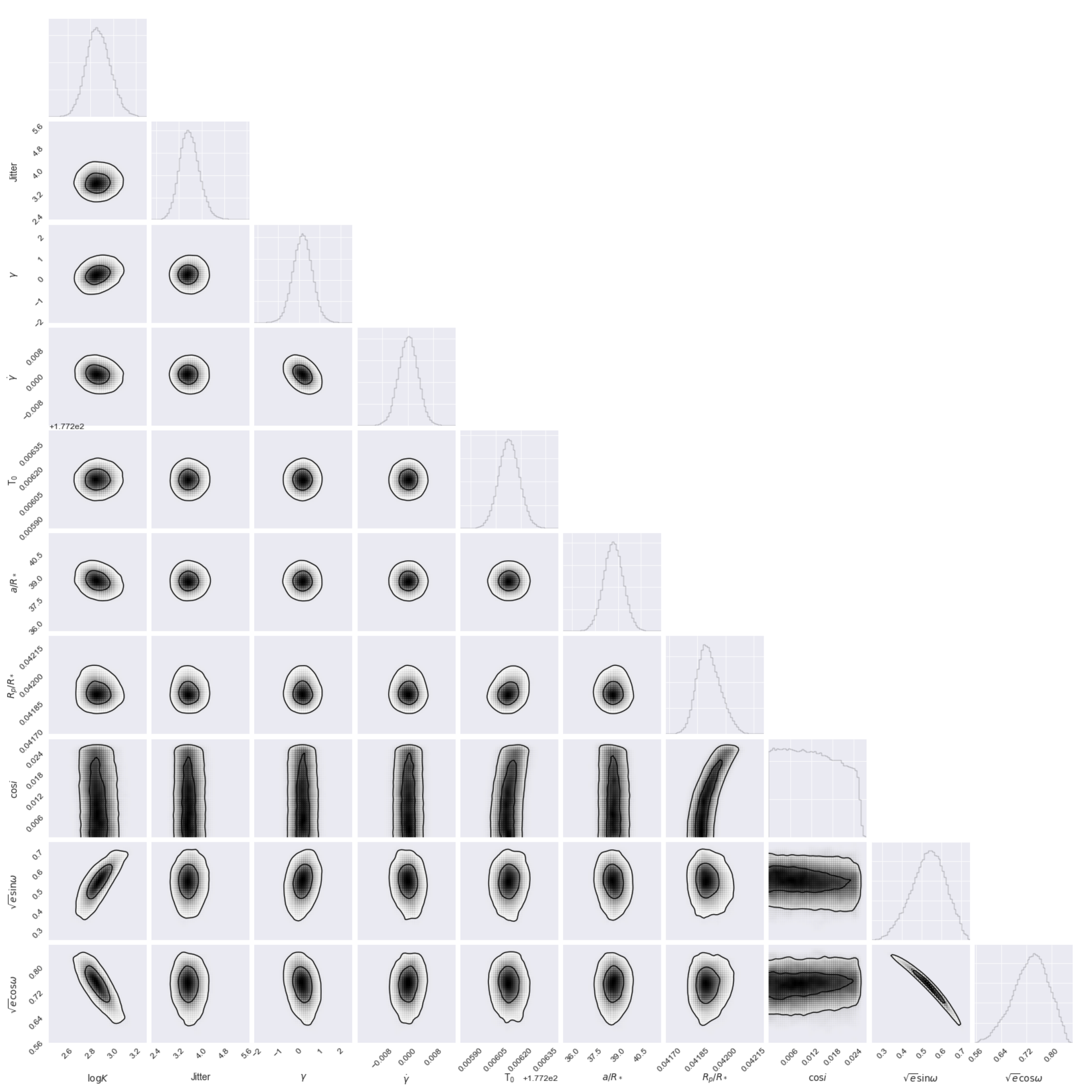}{1.0\textwidth}{}}
\caption{Joint constraints from the combined RV/transit model (Section~\ref{ssec:joint}).}
\end{figure*}

\bibliography{revtex-custom,manuscript}
\end{document}

%% file: all_params.tex
\usepackage{xstring} 

\newcommand{\koithreesixtysevenstar}[1]{%
  \IfEqCase{#1}{%
    {teff}{ 5735_{-67}^{+63}} 
    {logg}{ 4.366_{-0.091}^{+0.085}}
    {fe}{ 0.19_{-0.04}^{+0.04}} 
    {vsini}{ 2.8_{-1.0}^{+1.0}}
    {mass}{ 1.025_{-0.036}^{+0.038}} 
    {radius}{ 1.096_{-0.098}^{+0.133}} 
    {agegyr}{ 6.31_{-2.9}^{+2.1}} 
    {bmag}{ 11.90 \pm 0.10} 
    {vmag}{ 11.64 \pm 0.13} 
    {jmag}{ 10.046 \pm 0.020} 
    {hmag}{ 9.715 \pm 0.020} 
    {kmag}{ 9.640 \pm 0.017} 
    {u}{ 0.4160} 
    {v}{ 0.2495} 
  }[\PackageError{tree}{Undefined option to tree: #1}{}]%
}%

\newcommand{\koithreesixtysevenrv}[1]{%
  \IfEqCase{#1}{%
    {P1}{ 31.578659} 
    {T01}{ 2455010.206} 
    {k}{ 18.3_{-3.3}^{+9.6}} 
    {jit}{ 3.60_{-0.32}^{+0.37}} 
    {ecc}{ 0.844_{-0.053}^{+0.061}} 
    {omega}{ 51.6_{-10.0}^{+7.6}} 
    {dvdt}{ 0.0_{-1.1}^{+1.1}} 
    {gamma}{ 0.26_{-0.49}^{+0.49}} 
    {mpsini}{ 49.8_{-4.8}^{+10.2}} 
    {a}{ 0.1972_{-0.0033}^{+0.0033}} 
    {rp}{ 5.01_{-0.53}^{+0.53}} 
    {dp}{ 2.25_{-0.64}^{+1.03}} 
    {flux}{ 30.0_{-6.1}^{+6.9}} 
    {tp}{651_{-36}^{+34}} 
    {cmf}{0.827_{-0.063}^{+0.060}} 
    {t_ecc}{8548} 
  }[\PackageError{tree}{Undefined option to tree: #1}{}]%
}%

\newcommand{\koithreesixtyseventransit}[1]{%
  \IfEqCase{#1}{%
    {P1}{ 31.578659} 
    {T01}{ 2455010.20610_{-0.00081}^{+0.00083}} 
    {a_r}{ 38.71_{-0.65}^{+0.66}} 
    {rp_r}{ 0.0450_{-0.0033}^{+0.0008}} 
    {inc}{ 86.89_{-0.56}^{+0.36}} 
    {ecc}{ 0.75_{-0.08}^{+0.11}} 
    {omega}{ 87_{-59}^{+56}} 
    {a}{ 0.197_{-0.021}^{+0.021}} 
    {rp}{ 5.30_{-0.59}^{+0.60}} 
    {b}{ 0.63_{-0.28}^{+0.14}} 
  }[\PackageError{tree}{Undefined option to tree: #1}{}]%
}%

\newcommand{\koithreesixtysevenjoint}[1]{%
  \IfEqCase{#1}{%
    {P1}{ 31.578659} 
    {k}{ 17.6_{-1.7}^{+2.1}} 
    {jit}{ 3.56_{-0.33}^{+0.36}} 
    {gamma}{ 0.17_{-0.45}^{+0.45}} 
    {dvdt}{ 0.1_{-1.1}^{+1.1}} 
    {T01}{ 2455010.206131_{-0.000061}^{+0.000062}} 
    {a_r}{ 38.70_{-0.63}^{+0.65}} 
    {rp_r}{ 0.041908_{-0.000057}^{+0.000069}} 
    {inc}{ 89.31_{-0.51}^{+0.47}} 
    {ecc}{ 0.836_{-0.012}^{+0.013}} 
    {omega}{ 53.9_{-6.4}^{+6.7}} 
    {mpsini}{ 48.6_{-3.8}^{+4.2}} 
    {mp}{ 48.6_{-3.8}^{+4.2}} 
    {a}{ 0.197_{-0.021}^{+0.021}} 
    {rp}{ 5.02_{-0.53}^{+0.53}} 
    {dp}{ 2.13_{-0.57}^{+0.87}} 
    {flux}{ 30.0_{-7.8}^{+10.6}} 
    {tp}{ 651_{-47}^{+51}} 
    {cmf}{ 0.824_{-0.060}^{+0.067}} 
    {t_ecc}{ 8297_{-0}^{+0}} 
    {b}{ 0.083_{-0.057}^{+0.065}} 
  }[\PackageError{tree}{Undefined option to tree: #1}{}]%
}%

\newcommand{\val}[2]{%
  \IfEqCase{#1}{%
    {KOI-367-star}{\koithreesixtysevenstar{#2}}%
    {KOI-367-rv}{\koithreesixtysevenrv{#2}}%
    {KOI-367-tr}{\koithreesixtyseventransit{#2}}%
    {KOI-367-jo}{\koithreesixtysevenjoint{#2}}%
  }[\PackageError{tree}{Undefined option to tree: #1}{}]%
}%

%% file: tab_rv_stub.tex
2457521.024708 & 11.15 & 1.99 & 0.165 & 0.0017 \\
2457556.928676 & 0.35 & 2.13 & 0.161 & 0.0016 \\
2457556.960864 & 3.06 & 2.76 & 0.168 & 0.0017 \\
2457562.055606 & 10.94 & 1.94 & 0.158 & 0.0016 \\
2457562.924973 & 6.60 & 1.77 & 0.158 & 0.0016 \\
2457571.036004 & 2.43 & 2.57 & 0.124 & 0.0012 \\
2457581.987078 & -6.13 & 1.90 & 0.162 & 0.0016 \\
2457582.025562 & 0.76 & 1.66 & 0.157 & 0.0016 \\